\documentclass[aps, pra, reprint, superscriptaddress]{revtex4-1}
\usepackage{graphicx}
\usepackage{amsmath}
\usepackage{amssymb}
\usepackage{dcolumn}
\usepackage{color}
\usepackage{bm}

\begin{document}
\title{Effective temperature of a superfluid flowing in a random potential}
\author{Taiki Haga}
\email[]{haga@cat.phys.s.u-tokyo.ac.jp}
\affiliation{Department of Physics, University of Tokyo, 7-3-1 Hongo, Bunkyo-ku, Tokyo 113-0033, Japan}
\author{Masahito Ueda}
\affiliation{Department of Physics, University of Tokyo, 7-3-1 Hongo, Bunkyo-ku, Tokyo 113-0033, Japan}
\affiliation{RIKEN Center for Emergent Matter Science, 2-1 Hirosawa, Wako-shi, Saitama 351-0198, Japan}
\affiliation{Institute for Physics of Intelligence, University of Tokyo, 7-3-1 Hongo, Bunkyo-ku, Tokyo 113-0033, Japan}
\date{\today}

\begin{abstract}
The spatial fluctuations of a superfluid flowing in a weak random potential are investigated.
We employ classical field theory to demonstrate that the disorder-averaged nonequilibrium second-order correlation of the order parameter at zero temperature is identical to the thermally averaged equilibrium counterpart of a uniform superfluid at an effective temperature.
The physics behind this equivalence is that scattering of a moving condensate by disorder has the same effect on the correlation function as equilibrium thermal excitations.
The correlation function exhibits an exponential decay in one dimension and a power-law decay in two dimensions.
We show that the effective temperature can be measured in an interference experiment of ultracold atomic gases.
\end{abstract}

\maketitle

\section{Introduction}
\label{sec:introduction}

The universality of thermodynamics and statistical mechanics is attributed to the fact that macroscopic states of equilibrium systems can be described by a few key parameters such as temperature and pressure.
It is the lack of such phenomenological parameters in nonequilibrium states that makes it so challenging to establish a general framework of nonequilibrium statistical mechanics.
However, for some classes of nonequilibrium systems the notion of {\it effective temperature} is known to allow an approximate thermodynamic description of nonequilibrium states \cite{Casas-Vazquez-03, Cugliandolo-11}, and it has successfully been applied to a wide variety of driven systems including granular matter \cite{D'Anna-03, Wang-08}, structural glasses \cite{Crisanti-03, Cugliandolo-97, Berthier-00, Berthier-02},  coarsening systems \cite{Calabrese-05, Corberi-07}, turbulence \cite{Hohenberg-89}, and driven dissipative exciton-polariton systems \cite{Sieberer-13, Tauber-14}.

With remarkable experimental progress in ultracold atomic gases, the nonequilibrium dynamics of Bose-Einstein condensates has attracted considerable attention \cite{Dalfovo-99, Morsch-06, Lewenstein-07, Bloch-08}.
A superfluid exhibits dissipationless and stationary flow of matter as long as the flow velocity is below a certain critical value.
The absence of dissipation distinguishes a superfluid state with a nonzero current from typical nonequilibrium steady states maintained by the balance between external driving and energy dissipation.
It is therefore of fundamental interest to investigate whether the notion of an effective temperature is viable in nonequilibrium superfluid systems.

In this study, we consider a superflow in a weak random potential.
The effect of disorder on superfluidity has been a long-standing problem in condensed matter physics \cite{Fisher-89, Huang-92, Singh-92, Giorgini-94, Freericks-96, Falco-09, Altman-10, Ristivojevic-12, Zuniga-15}.
In the ground state, weak disorder does not affect the global phase coherence except for a small depletion of the condensate fraction \cite{Huang-92, Giorgini-94}.
The stability of a superflow in a random potential has also been investigated theoretically \cite{Paul-07, Albert-08, Paul-09, Albert-10} and experimentally \cite{Lye-05, Clement-05, Chen-08, Dries-10, Tanzi-13}.
In this work, we focus on the spatial phase fluctuations of the superfluid order parameter in a stationary flow.
Within classical field theory, we demonstrate that the disorder-averaged nonequilibrium second-order correlation of the order parameter at zero temperature is identical to the thermally-averaged equilibrium counterpart of a uniform superfluid at an effective temperature.
In particular, the disorder-averaged correlation exhibits an exponential decay in one dimension and a power-law decay in two dimensions.
We argue that scattering of a moving condensate by disorder has the same effect as thermal excitations, and the superfluid flowing in a random potential can be identified with a uniform system at thermal equilibrium with an effective temperature.
The effective temperature is shown to be proportional to the square of disorder strength and that of the flow velocity in a weak disorder and small velocity regime, and diverges as the flow velocity approaches the sound velocity of the condensate.

The decay behavior of the disorder-averaged correlation in one and two dimensions is reminiscent of the Hohenberg--Mermin--Wagner theorem for a system with continuous symmetry, which states that the thermally-averaged correlation of the order parameter decays in one and two dimensions \cite{Mermin-66, Hohenberg-67, Coleman-73}.
It is of fundamental importance in nonequilibrium statistical physics to understand how and when nonequilibrium driving destroys an ordered phase that is stable in equilibrium \cite{Janot-13}.
Our study offers a general mechanism responsible for the breakdown of off-diagonal long-range order of a superfluid due to the interplay between nonequilibrium current and disorder.

This paper is organized as follows.
In Sec.~\ref{sec:model}, we introduce a generic model for disordered Bose systems.
We employ the classical field approximation, in which the dynamics of a bosonic field is described by a c-number field obeying a nonlinear Schr\"odinger equation.
In Sec.~\ref{sec:effective_temperature}, we calculate the correlation function of the superfluid order parameter in the presence of a flow at zero temperature and show that it is identical to that of a thermal-equilibrium uniform superfluid at an effective temperature.
We discuss the underlying mechanism for the emergence of such thermal behavior.
In Sec.~\ref{sec:Bose_gas_in_random_poential}, we derive an explicit expression of the correlation function of a Bose gas flowing in a random potential to the leading order in the disorder strength.
We estimate the effective temperature by using typical parameters for ultracold atomic gases.
In Sec.~\ref{sec:numerical_simulations}, we numerically calculate the disorder-averaged correlation function to confirm the perturbative results obtained in Sec.~\ref{sec:Bose_gas_in_random_poential}.
We find that our qualitative predictions for the correlation function still hold for a moderately strong disorder.
In Sec.~\ref{sec:interference_and_correlation}, we propose an interference experiment to measure the disorder-averaged correlation for a Bose gas flowing in a random potential.
Finally, Sec.~\ref{sec:concluding_remarks} is devoted to conclusions and outlook for future study.

\section{Model}
\label{sec:model}

In the classical field theory, bosonic field operators in a quantum Hamiltonian are replaced by a classical field $\Phi$.
We consider a single-component superfluid described by the following Hamiltonian in spatial dimension $D$ subject to the periodic boundary conditions with period $L$ in all directions:
\begin{equation}
H[\Phi] = \int d^D \mathbf{r} \biggl[ \frac{1}{2} Z(n;\lambda) |\nabla \Phi|^2 + U(n;\kappa) \biggr],
\label{General_Hamiltonian}
\end{equation}
where $Z(n;\lambda) \: (>0)$ and $U(n;\kappa)$ are analytic functions of the density $n := |\Phi|^2$ and randomness parameters $\lambda$ and $\kappa$. 
Explicit forms of these functions are not needed for the general discussion in the next section.
The time-independent and spatially fluctuating parameters $\lambda(\mathbf{r})$ and $\kappa(\mathbf{r})$ characterize spatially irregular structures of the system, such as porous media for superfluid helium \cite{Reppy-92} and optical speckle patterns for ultracold atomic gases \cite{Lye-05, Clement-05, Chen-08, Dries-10}.
We assume that $\overline{\lambda(\mathbf{r})}=\overline{\kappa(\mathbf{r})}=0$, where the overline denotes the disorder average, and that the spatial correlations of $\lambda(\mathbf{r})$ and $\kappa(\mathbf{r})$ decay exponentially with the distance.
The time evolution of $\Phi$ is described by
\begin{equation}
i\hbar \frac{\partial \Phi}{\partial t} = \frac{\delta H[\Phi]}{\delta \Phi^*}.
\label{General_EOM}
\end{equation}

The classical field theory is valid when the quantum depletion due to interactions is negligible and almost all bosons occupy a single-particle state.
However, for a quasi-one-dimensional case, in which a weakly interacting Bose gas is tightly confined in a cylindrical trap, special care is needed because quantum fluctuations lead to an algebraic decay of phase coherence.
We define a phase coherence length $R_{\varphi}$ characterizing the length scale of phase fluctuations due to quantum effects.
In terms of the healing length $\xi_{\mathrm{h}}$, the three-dimensional scattering length $a$, the transverse confinement length $l_0$, and the density $n$ per unit longitudinal length, the phase coherence length is given by $R_{\varphi}\sim\xi_{\mathrm{h}}\exp(\pi l_0 \sqrt{n/2a})$ \cite{Paul-09, Petrov-00}.
Thus, when $(a/l_0)^2 \ll na$, $R_{\varphi}$ is exponentially larger than $\xi_{\mathrm{h}}$.
The classical field theory is justified when the system size $L$ is much smaller than $R_{\varphi}$.
Since $a/l_0$ is of the order of $10^{-3}$ in typical experiments of ultracold atomic gases, there exists a sufficiently broad range of $L$ in which quantum fluctuations are negligible \cite{Leboeuf-01}.

A typical example of systems described by Eq.~(\ref{General_Hamiltonian}) is a weakly interacting Bose gas in a weak random potential, for which the Hamiltonian is given by
\begin{equation}
H[\Phi] = \int d^D \mathbf{r} \biggl[ \frac{\hbar^2}{2m} |\nabla \Phi(\mathbf{r})|^2 + V(\mathbf{r}) n(\mathbf{r}) + \frac{g}{2} n(\mathbf{r})^2 \biggr],
\label{Hamiltonian}
\end{equation}
where $m$ is the mass of an atom and $g\:(>0)$ is the strength of a repulsive interaction.
A zero-mean random potential $V(\mathbf{r})$ satisfies 
\begin{equation}
\overline{V(\mathbf{r})V(\mathbf{r}')} = C_{\mathrm{R}}(\mathbf{r}-\mathbf{r}'),
\end{equation}
where $C_{\mathrm{R}}(\mathbf{r})$ is short-ranged with a characteristic length scale $\xi_{\mathrm{R}}$; for example, $C_{\mathrm{R}}(\mathbf{r}) = V_0^2 \exp( -|\mathbf{r}|^2/2\xi_{\mathrm{R}}^2 )$ \cite{Piraud-13}.
We assume that the amplitude of the random potential $V_0 := C_{\mathrm{R}}(\mathbf{0})^{1/2}$ is much smaller than the interaction energy $gn$.
Equations (\ref{General_EOM}) and (\ref{Hamiltonian}) lead to the time-dependent Gross-Pitaevskii (GP) equation:
\begin{equation}
i\hbar \frac{\partial \Phi(\mathbf{r})}{\partial t} = -\frac{\hbar^2}{2m} \nabla^2 \Phi(\mathbf{r}) + [V(\mathbf{r}) + g n(\mathbf{r})]\Phi(\mathbf{r}).
\label{time_dependent_GP_eq}
\end{equation}

\section{Effective temperature}
\label{sec:effective_temperature}

We investigate the spatial fluctuations of the order parameter $\Phi$ described by the general Hamiltonian (\ref{General_Hamiltonian}) in the presence of a flow.
From Eqs.~(\ref{General_Hamiltonian}) and (\ref{General_EOM}), the continuity equation for the particle density reads
\begin{equation}
\partial_t n = - \nabla \cdot \mathbf{j},
\label{continuity_equation}
\end{equation}
where the current density $\mathbf{j}$ is given by 
\begin{equation}
\mathbf{j} = - \frac{i}{2} Z(n;\lambda) (\Phi^* \nabla \Phi - \Phi \nabla \Phi^*).
\end{equation}
We focus on a stationary state: $i\hbar \partial_t \Phi=\mu \Phi$, where $\mu$ is the chemical potential.
If the flow velocity is smaller than the critical velocity, a stable stationary solution of Eq.~(\ref{General_EOM}) exists.
We denote such a solution as 
\begin{equation}
\Phi(\mathbf{r}) = \sqrt{n(\mathbf{r})} e^{i \mathbf{K} \cdot \mathbf{r} + i\varphi(\mathbf{r})},
\label{Phi_ansatz}
\end{equation}
where $\mathbf{K}$ is the average momentum and $\varphi(\mathbf{r})$ describes phase fluctuations.
We assume that $\overline{\varphi(\mathbf{r})}=0$ and typical variations of $\varphi$ are small for sufficiently small $|\lambda|$ and $|\kappa|$.
From the continuity equation (\ref{continuity_equation}), we have
\begin{equation}
\nabla \cdot [Z(n;\lambda) n (\nabla \varphi + \mathbf{K})] = 0.
\label{continuity_equation_stationary}
\end{equation}

For a given realization of disorder, $n$ and $Z$ can be split into their spatial averages and the deviations therefrom: $n(\mathbf{r}) = \bar{n} + \delta n(\mathbf{r})$ and $Z(n(\mathbf{r});\lambda(\mathbf{r})) = \bar{Z} + \delta Z(\mathbf{r})$, where $\bar{n} := L^{-D} \int d^D \mathbf{r} n(\mathbf{r})$ and $\bar{Z} := L^{-D} \int d^D \mathbf{r} Z(n(\mathbf{r});\lambda(\mathbf{r}))$.
To the leading order, Eq.~(\ref{continuity_equation_stationary}) gives
\begin{equation}
\bar{n} \nabla^2 \varphi(\mathbf{r}) + \mathbf{K} \cdot \nabla \delta \tilde{n}(\mathbf{r}) = 0,
\label{continuity_equation_leading_order}
\end{equation}
where $\delta \tilde{n}(\mathbf{r}) := \delta n(\mathbf{r})+\bar{n} \delta Z(\mathbf{r})/\bar{Z}$.
By introducing the Fourier transform of $\varphi(\mathbf{r})$,
\begin{equation}
\varphi_{\mathbf{q}} = L^{-D/2} \int d^D\mathbf{r} \varphi(\mathbf{r}) e^{-i \mathbf{q} \cdot \mathbf{r}},
\end{equation}
where $\mathbf{q} = 2\pi\mathbf{n}/L$, $(\mathbf{n} \in \mathbb{Z}^D)$, we have
\begin{equation}
\overline{|\varphi_{\mathbf{q}}|^2} = \frac{(\mathbf{K} \cdot \mathbf{q})^2}{|\mathbf{q}|^4} \bar{n}^{-2} \overline{|\delta \tilde{n}_{\mathbf{q}}|^2}, \:\:\: (\mathbf{q} \neq \mathbf{0}).
\label{phase_fluctuation}
\end{equation}

Next, we consider the correlation function of density fluctuations,
\begin{equation}
g_n(\mathbf{r}-\mathbf{r}') := \overline{\delta \tilde{n}(\mathbf{r}) \delta \tilde{n}(\mathbf{r}')},
\end{equation}
whose Fourier transform gives 
\begin{equation}
\overline{|\delta \tilde{n}_{\mathbf{q}}|^2} = \int d^D \mathbf{r} g_n(\mathbf{r}) e^{-i\mathbf{q} \cdot \mathbf{r}}.
\end{equation}
We assume that $g_n(\mathbf{r})$ consists of an exponentially decaying part and a negative offset, the latter of which is inversely proportional to the volume: 
\begin{equation}
g_n(\mathbf{r}) \simeq  \sigma_n^2 e^{-|\mathbf{r}|/\xi_n} - \kappa L^{-D}, 
\end{equation}
where $\sigma_n^2 = \overline{\delta \tilde{n}(\mathbf{0})^2} + \kappa L^{-D}$ is the amplitude of the density fluctuations and $\xi_n$ is the correlation length.
The constant $\kappa>0$ is determined from the condition 
\begin{equation}
\int d^D \mathbf{r} g_n(\mathbf{r}) = 0, 
\end{equation}
which follows from the definition of $\delta \tilde{n}(\mathbf{r})$.
Because $\xi_n$ can depend on the relative angle between $\mathbf{r}$ and $\mathbf{K}$, it is convenient to define $\tilde{\xi}_n$ as 
\begin{equation}
\int d^D \mathbf{r} (g_n(\mathbf{r}) + \kappa L^{-D}) = \sigma_n^2 \tilde{\xi}_n^D.
\end{equation}
Then, $\overline{|\delta \tilde{n}_{\mathbf{q}}|^2}$ converges to $\sigma_n^2 \tilde{\xi}_n^D$ as $\mathbf{q}$ vanishes, giving
\begin{equation}
\overline{|\varphi_{\mathbf{q}}|^2} \sim \frac{(\mathbf{K} \cdot \mathbf{q})^2}{|\mathbf{q}|^4} \bar{n}^{-2} \sigma_n^2 \tilde{\xi}_n^D
\label{phase_fluctuation_small_q}
\end{equation}
for small $\mathbf{q}$.
Remarkably, $|\varphi_{\mathbf{q}}|^2$ diverges in the long-wavelength limit as $\propto |\mathbf{q}|^{-2}$, which implies that the amplitude of phase fluctuations behaves as
\begin{eqnarray}
\sigma_{\varphi}^2 := L^{-D} \int d^D\mathbf{r} \overline{\varphi(\mathbf{r})^2} \sim \left\{ \begin{array}{ll}
L, & (D=1); \\
\ln L, & (D=2). \\
\end{array} \right.
\end{eqnarray}

We define the disorder-averaged correlation function of the order parameter by 
\begin{equation}
C(\mathbf{r} - \mathbf{r}') := e^{-i \mathbf{K} \cdot (\mathbf{r}-\mathbf{r}')} \overline{\Phi(\mathbf{r}) \Phi^*(\mathbf{r}')}, 
\label{def_C}
\end{equation}
where $\Phi(\mathbf{r})$ is a stationary solution of Eq.~(\ref{General_EOM}).
The phase factor in Eq.~(\ref{def_C}) has been introduced to compensate the mean flow $\mathbf{K}$ in Eq.~(\ref{Phi_ansatz}).
Because $g_n(\mathbf{r})$ decays exponentially, Eq.~(\ref{def_C}) is approximated as
\begin{equation}
C(\mathbf{r}) \simeq \bar{n} \overline{e^{i(\varphi(\mathbf{r})- \varphi(\mathbf{0}))}},
\end{equation}
for $|\mathbf{r}| \gg \tilde{\xi}_n$.
In terms of the mean square relative displacement of the U(1) phase
\begin{equation}
B(\mathbf{r}) := \overline{(\varphi(\mathbf{r}) - \varphi(\mathbf{0}))^2}, 
\end{equation}
the correlation function is rewritten as 
\begin{equation}
C(\mathbf{r}) \simeq \bar{n} \exp \left[-\frac{1}{2} B(\mathbf{r})\right],
\end{equation}
where we have retained only the second cumulant of the phase fluctuations.
From Eqs.~(\ref{phase_fluctuation}) and (\ref{phase_fluctuation_small_q}), the asymptotic behavior of the mean square relative displacement of the U(1) phase for long distance $|\mathbf{r}| \gg \tilde{\xi}_n$ can be calculated as 
\begin{eqnarray}
\hspace{-1em} B(\mathbf{r}) \sim \left\{ \begin{array}{ll}
K^2 \bar{n}^{-2} \sigma_n^2 \tilde{\xi}_n |r|, & (D=1); \\
(2\pi)^{-1} K^2 \bar{n}^{-2} \sigma_n^2 \tilde{\xi}_n^2 \ln (|\mathbf{r}|/\tilde{\xi}_n), & (D=2), \\
\end{array} \right.
\end{eqnarray}
where details of the calculation are presented in Appendix \ref{appendix:mean_square_relative_displacement}.
Thus, we have
\begin{eqnarray}
C(\mathbf{r}) \sim \left\{ \begin{array}{ll}
e^{-|r|/l_{\mathrm{c}}}, & (D=1); \\
(|\mathbf{r}|/\tilde{\xi}_n)^{-\eta}, & (D=2), \\
\end{array} \right.
\label{C_asymptotic}
\end{eqnarray}
where the inverse correlation length $l_{\mathrm{c}}^{-1}$ and the exponent $\eta$ are given by
\begin{equation}
l_{\mathrm{c}}^{-1} = \frac{K^2  \sigma_n^2 \tilde{\xi}_n}{2\bar{n}^2},
\label{l_c}
\end{equation}
\begin{equation}
\eta = \frac{K^2 \sigma_n^2 \tilde{\xi}_n^2}{4\pi \bar{n}^2 }.
\label{eta}
\end{equation}
In two dimensions, although $C(\mathbf{r})$ is anisotropic, the exponent $\eta$ is independent of the direction of $\mathbf{r}$.

The long-distance behavior of the correlation function in Eq.~(\ref{C_asymptotic}) is the same as that of a uniform Bose gas at thermal equilibrium with temperature $T$.
In such a case, the inverse correlation length in one dimension and the exponent in two dimensions are given by
\begin{equation}
l_{\mathrm{c}, \mathrm{eq}}^{-1} = \frac{m k_{\mathrm{B}}T}{\bar{n} \hbar^2}, 
\label{l_c_uniform}
\end{equation}
\begin{equation}
\eta_{\mathrm{eq}} = \frac{m k_{\mathrm{B}}T}{2\pi \bar{n} \hbar^2}.
\label{eta_uniform}
\end{equation}
Comparing Eqs.~(\ref{l_c}) and (\ref{eta}) with Eqs.~(\ref{l_c_uniform}) and (\ref{eta_uniform}), we are led to introduce an effective temperature
\begin{equation}
k_{\mathrm{B}} T_{\mathrm{eff}} = \frac{\hbar^2 K^2  \sigma_n^2 \tilde{\xi}_n^D}{2 m \bar{n}}.
\label{T_eff}
\end{equation}
This effective temperature can be rewritten as 
\begin{equation}
k_{\mathrm{B}} T_{\mathrm{eff}} = \left( \frac{\hbar^2 K^2}{2m} \right) \times \left( \bar{n}\tilde{\xi}_n^D \right) \times \left( \frac{\sigma_n^2}{\bar{n}^2} \right), 
\label{T_eff_2}
\end{equation}
where the first, second, and third terms represent the kinetic energy of the condensed atoms, the number of atoms within the correlation length, and the amplitude of the density fluctuations, respectively.
The last term $\sigma_n^2/\bar{n}^2$ is interpreted to be the scattering probability of the condensed atom by the random medium.
For example, suppose that a Bose gas flows in a weak random potential $V(\mathbf{r})$.
Fermi's golden rule implies that the transition rate between plane wave states with momenta $\mathbf{k}$ and $\mathbf{k}'$ is proportional to $|\langle \mathbf{k}'|V(\mathbf{r})|\mathbf{k} \rangle|^2 = \int d^D \mathbf{r} \overline{V(\mathbf{r}) V(\mathbf{0})} e^{i (\mathbf{k}-\mathbf{k}') \cdot \mathbf{r}}$.
Because the amplitude of the density fluctuations is proportional to that of the random potential, $\sigma_n^2/\bar{n}^2$ is proportional to the number of atoms that are scattered out of the condensate.
Thus, Eq.~(\ref{T_eff_2}) implies that the scattering process due to the random potential is equivalent to virtual thermal excitations.
This process does not lead to an actual heating of the system because neither injection nor dissipation of energy is involved.

\section{Bose gas flowing in a random potential}
\label{sec:Bose_gas_in_random_poential}

In this section, we explicitly calculate the correlation function $C(\mathbf{r})$ for a weakly interacting Bose gas flowing in a weak random potential.
The dynamics is described by Eq.~(\ref{time_dependent_GP_eq}), and thus, the stationary state satisfies the time-independent GP equation:
\begin{equation}
-\frac{\hbar^2}{2m} \nabla^2 \Phi(\mathbf{r}) + [V(\mathbf{r}) + g n(\mathbf{r})]\Phi(\mathbf{r}) = \mu \Phi(\mathbf{r}).
\label{GP_eq}
\end{equation}
For a given mean density $\bar{n}$, the chemical potential $\mu$ is determined from the condition 
\begin{eqnarray}
L^{-D} \int d\mathbf{r} n(\mathbf{r}) = \bar{n},
\label{density_constraint}
\end{eqnarray}
where $n(\mathbf{r})$ is the solution to Eq.~(\ref{GP_eq}).
We consider a stationary state given by Eq.~(\ref{Phi_ansatz}).
The density $n(\mathbf{r})$, the phase $\varphi(\mathbf{r})$, and the chemical potential $\mu$ can be expanded with respect to the disorder strength as 
\begin{equation}
\begin{split}
n(\mathbf{r}) &= \bar{n} + n^{(1)}(\mathbf{r}) + n^{(2)}(\mathbf{r}) + \cdots, \\
\varphi(\mathbf{r}) &= \varphi^{(1)}(\mathbf{r}) + \varphi^{(2)}(\mathbf{r}) + \cdots, \\
\mu &= \mu^{(0)} + \mu^{(1)} + \mu^{(2)} + \cdots,
\end{split}
\label{expansion}
\end{equation}
where $n^{(s)}(\mathbf{r})$, $\varphi^{(s)}(\mathbf{r})$, and $\mu^{(s)}$ $(s=1,2,\cdots)$ are of the order of $O(V^s)$.
From Eqs.~(\ref{GP_eq}) and (\ref{density_constraint}), we have
\begin{eqnarray}
\mu^{(0)} = g\bar{n} + \frac{\hbar^2 K^2}{2m}, \:\:\:\:\: \mu^{(1)} = 0.
\end{eqnarray}

Substituting Eq.~(\ref{Phi_ansatz}) into Eq.~(\ref{GP_eq}), we have
\begin{widetext}
\begin{eqnarray}
-\frac{\hbar^2}{2m} \bigl[ \nabla^2 \sqrt{n(\mathbf{r})} - 2\sqrt{n(\mathbf{r})} (\mathbf{K} \cdot \nabla) \varphi(\mathbf{r}) - \sqrt{n(\mathbf{r})} |\nabla \varphi(\mathbf{r})|^2 - K^2 \sqrt{n(\mathbf{r})} \bigr] + (gn(\mathbf{r}) + V(\mathbf{r}) - \mu) \sqrt{n(\mathbf{r})} = 0,
\label{density_eq}
\end{eqnarray}
\begin{eqnarray}
\nabla \cdot [n(\mathbf{r}) (\nabla \varphi(\mathbf{r}) + \mathbf{K})] = 0,
\label{continuity_equation_stationary_RP}
\end{eqnarray}
\end{widetext}
where the second equation is the equation of continuity.
Inserting Eq.~(\ref{expansion}) into (\ref{density_eq}) and (\ref{continuity_equation_stationary_RP}) and keeping only the leading-order terms, we have
\begin{equation}
\frac{\hbar^2}{2m} \bigl[ |\mathbf{q}|^2 n^{(1)}_{\mathbf{q}} + 4\bar{n} K iq_x \varphi^{(1)}_{\mathbf{q}} \bigr] + 2g\bar{n} n^{(1)}_{\mathbf{q}} + 2\bar{n} V_{\mathbf{q}} = 0,
\label{n_1_eq_Fourier}
\end{equation}
\begin{equation}
-\bar{n} |\mathbf{q}|^2 \varphi^{(1)}_{\mathbf{q}} + K iq_x n^{(1)}_{\mathbf{q}} = 0,
\label{phi_1_eq_Fourier}
\end{equation}
where $\mathbf{K}$ is assumed to be parallel to the positive $x$-direction, i.e., $\mathbf{K}=K \mathbf{e}_x$ with $K>0$.
In terms of the flow velocity $v=\hbar K/m$, the sound velocity $c=\sqrt{g\bar{n}/m}$, and the healing length $\xi_{\mathrm{h}}=\hbar/\sqrt{2mg\bar{n}}$, $n^{(1)}(\mathbf{r})$ and $\varphi^{(1)}(\mathbf{r})$ are written as
\begin{equation}
n^{(1)}_{\mathbf{q}} = - \frac{\bar{n} |\mathbf{q}|^2}{mc^2 [|\mathbf{q}|^2 - (v/c)^2 q_x^2 + \xi_{\mathrm{h}}^2 |\mathbf{q}|^4/2]}  V_{\mathbf{q}},
\label{A_1}
\end{equation}
\begin{equation}
\varphi^{(1)}_{\mathbf{q}} = - \frac{i v q_x }{\hbar c^2 [|\mathbf{q}|^2 - (v/c)^2 q_x^2 + \xi_{\mathrm{h}}^2 |\mathbf{q}|^4/2]}  V_{\mathbf{q}}.
\label{phi_1}
\end{equation}
In a similar manner as in Sec.~\ref{sec:effective_temperature} and Appendix \ref{appendix:mean_square_relative_displacement}, the inverse correlation length $l_{\mathrm{c}}^{-1}$ in one dimension and the exponent $\eta$ in two dimensions can be calculated as
\begin{equation}
l_{\mathrm{c}}^{-1} = \frac{v^2 \tilde{C}_{\mathrm{R}}(0)}{2\hbar^2(c^2 - v^2)^2},
\label{l_c_RP}
\end{equation}
\begin{equation}
\eta = \frac{ v^2 \tilde{C}_{\mathrm{R}}(\mathbf{0})}{4\pi \hbar^2 c(c^2 - v^2)^{3/2}},
\label{eta_RP}
\end{equation}
where $\tilde{C}_{\mathrm{R}}(\mathbf{q}) = \int d^D\mathbf{r} \: C_{\mathrm{R}}(\mathbf{r}) e^{-i \mathbf{q} \cdot \mathbf{r}}$.
It should be noted that the corresponding effective temperature diverges as the flow velocity $v$ approaches the sound velocity $c$.

Let us estimate the typical values of $l_{\mathrm{c}}$, $\eta$, and $T_{\mathrm{eff}}$ by employing experimental parameters in Ref.~\cite{Lye-05}. 
We take the correlation length and the amplitude of the random potential as $\xi_{\mathrm{R}} = 10 \mathrm{\mu m}$ and $V_{\mathrm{R}} = 100 \mathrm{Hz} \times h$, which is much smaller than the chemical potential $\mu = gn = 1 \mathrm{kHz} \times h$.
The disorder correlator at zero wavenumber is given by $\tilde{C}_{\mathrm{R}}(0) \simeq V_0^2 \xi_{\mathrm{R}}$ in one dimension and $\tilde{C}_{\mathrm{R}}(\mathbf{0}) \simeq V_0^2 \xi_{\mathrm{R}}^2$ in two dimensions.
The sound velocity is estimated as $c=\sqrt{gn/m}=2.2 \times 10^{-3} \mathrm{m s^{-1}}$ for $\mathrm{Rb}$ atoms.
For a flow velocity $v=c/3$, from Eqs.~(\ref{l_c_RP}) and (\ref{eta_RP}) we obtain $l_{\mathrm{c}}=17\mathrm{\mu m}$ in one dimension and $\eta=0.086$ in two dimensions.
We assume that the three-dimensional particle density and the confinement length are given by $n_{\mathrm{3D}}=10^{14}\mathrm{cm}^{-3}$ and $l_0=3\mathrm{\mu m}$, respectively.
Then,  in one dimension the density per unit length reads $n_{\mathrm{1D}} = \pi(l_0/2)^2 n_{\mathrm{3D}} = 7.1\times10^6 \mathrm{cm}^{-1}$, and in two dimensions the density per unit area reads  $n_{\mathrm{2D}} = l_0 n_{\mathrm{3D}} = 3\times 10^{10} \mathrm{cm}^{-2}$.
Thus, the effective temperature is estimated as
\begin{equation}
T_{\mathrm{eff}}^{(\mathrm{1D})} = \frac{n_{\mathrm{1D}} \hbar^2 l_{\mathrm{c}}^{-1}}{m k_{\mathrm{B}}} \simeq 240 \mathrm{nK}
\label{T_eff_1D}
\end{equation}
in one dimension and 
\begin{equation}
T_{\mathrm{eff}}^{(\mathrm{2D})} = \frac{2 \pi n_{\mathrm{2D}} \hbar^2 \eta}{m k_{\mathrm{B}}} \simeq 950 \mathrm{nK}
\label{T_eff_2D}
\end{equation}
in two dimensions.
To measure $T_{\mathrm{eff}}$, the thermodynamic temperature of condensates has to be sufficiently lower than the values of Eqs.~(\ref{T_eff_1D}) and (\ref{T_eff_2D}).
This requirement is well met experimentally in ultracold atoms.

\section{Numerical simulations}
\label{sec:numerical_simulations}

Let us calculate $C(\mathbf{r})$ by numerically solving the GP equation.
We consider the Bose--Hubbard model in one- and two-dimensional square lattices.
If the number of atoms per site is sufficiently large, the classical field theory is applicable \cite{Polkovnikov-02}.
The discrete GP equation reads
\begin{eqnarray}
-J \sum_{k \in \mathcal{N}_j} \Phi_{k} + (U_{j} + g n_j) \Phi_{j} = \mu \Phi_{j},
\label{stationary_equation_lattice}
\end{eqnarray}
where $\mathcal{N}_j$ denotes the set of the nearest neighbor sites of $j$ and $n_j=|\Phi_{j}|^2$ is the particle-number density at site $j$. 
The on-site potential $U_{j}$ is randomly chosen from a uniform distribution over the interval $[-W,W]$.

We seek for the stationary solution of the form 
\begin{eqnarray}
\Phi_{j}=\sqrt{n_j} e^{i \mathbf{K} \cdot \mathbf{R}_j + i\varphi_j},
\end{eqnarray}
where $\mathbf{R}_j=(R_j^x, R_j^y)$ is the coordinate of site $j$, and $\mathbf{K}$ is assumed to be parallel to the positive $x$ direction.
To obtain such a solution numerically, we consider the following time-dependent GP equation:
\begin{eqnarray}
\hspace{-1em} i \frac{d\Phi_{j}}{dt} = -J \sum_{k \in \mathcal{N}_j} e^{i\theta(t) (R_k^x-R_j^x)} \Phi_{k} + (U_{j} + g n_{j}) \Phi_{j}.
\label{time_dependent_GP_lattice}
\end{eqnarray}
At an initial time $t=0$, we start from the ground state of the Hamiltonian, which is obtained from the solution of the imaginary time GP equation.
Then, the phase parameter $\theta(t)$ is gradually increased up to a given value $K$ during a time interval $\tau$: $\theta(t)=Kt/\tau$.
If $K$ is smaller than the critical momentum $K_{\mathrm{c}}$, one reaches a stationary solution at $t=\tau$.
The time interval $\tau$ is chosen to be sufficiently long so that the solution is well converged.
Above $K_{\mathrm{c}}$, there exists no stationary solution that is continuously connected to the ground state of the Hamiltonian.
Equation (\ref{time_dependent_GP_lattice}) is numerically solved by the Runge-Kutta method with a discretized time step $dt=0.1$.
In the following, we set $J=g=1$ and $\bar{n} = L^{-D} \sum_{j} |\Phi_{j}|^2 =1$.
To compare numerical results for the lattice system with the perturbative ones for the continuous system discussed in Sec.~\ref{sec:Bose_gas_in_random_poential}, the following replacements need to be made:
\begin{equation}
\begin{split}
\hbar^2 |\mathbf{q}|^2/2m \: &\to \: 2J (2 - \cos q_x - \cos q_y), \\
m v/\hbar \: &\to \: K, \\
\hbar c \: &\to \: (2Jg\bar{n})^{1/2}.
\end{split}
\end{equation}

\begin{figure}
 \centering
 \includegraphics[width=0.4\textwidth]{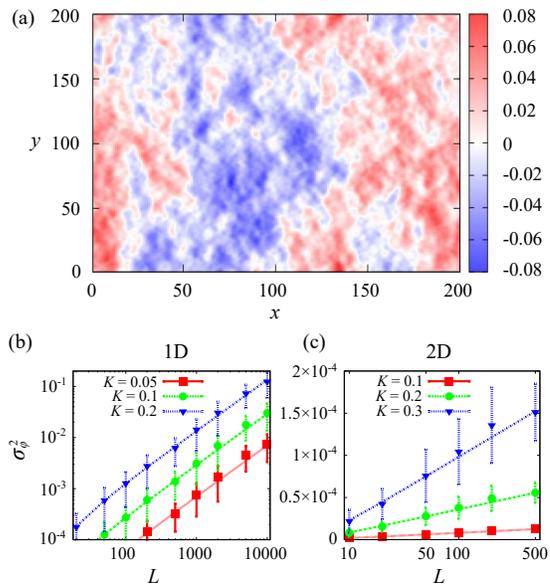}
 \caption{(a) Phase fluctuation $\varphi(x,y)$ of a stationary state in two dimensions with disorder strength $W=0.1$, flow momentum $K=0.4$, and system size $L=200$.
 (b), (c) Variance of phase fluctuations, $\sigma_{\varphi}^2$, as a function of the system size $L$ for different values of $K$ with $W=0.1$ in (b) one dimension and (c) two dimensions.
 The error bars show one standard deviation for different realizations of the random potential.
 Figures (b) and (c) are presented in double-log and semi-log plots, respectively. 
 The dashed lines show the perturbative results calculated from Eq.~(\ref{phi_1}).}
 \label{fig-phase-fluctuation}
\end{figure}

Figure \ref{fig-phase-fluctuation} (a) shows the spatial distribution of $\varphi_j$ for a stationary state in two dimensions.
Note that the phase fluctuations are more strongly correlated for the direction perpendicular to the flow velocity.
Figures \ref{fig-phase-fluctuation} (b) and (c) show $\sigma_{\varphi}^2 := L^{-D} \sum_{j} \overline{\varphi_{j}^2}$ as a function of the system size $L$ for one and two dimensions.
The disorder strength is $W=0.1$, which is much smaller than $g\bar{n}=1$.
For the one-dimensional case (b), the time interval $\tau$ is set to $10^4$.
The disorder average is taken over 100, 50, 20, and 10 realizations of the random potential for different system sizes $L=20-1000$, $L=2000$, $L=5000$, and $L=10000$, respectively.
For the two-dimensional case (c), the time interval $\tau$ is set to $2 \times 10^3$.
The disorder average is taken over 100, 50, 20, and 10 realizations of the random potential for different system sizes $L=10-50$, $L=100$, $L=200$, and $L=500$, respectively.
We find that $\sigma_{\varphi}^2 \propto L$ in one dimension, and $\sigma_{\varphi}^2 \propto \ln L$ in two dimensions.
The dashed lines in (b) and (c) represent the perturbative results obtained from Eq.~(\ref{phi_1}).
The agreement between the numerical and analytical results is excellent.

\begin{figure}
 \centering
 \includegraphics[width=0.45\textwidth]{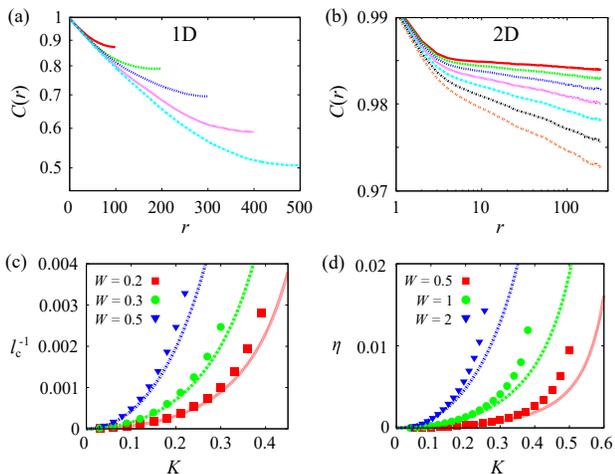}
 \caption{(a) Correlation function $C(r)$ in one dimension with disorder strength $W=0.5$ and flow momentum $K=0.2$ for system size $L=200$, $400$, $600$, $800$, $1000$ from top to bottom.
 The abscissa and ordinate are shown in linear and log scales, respectively.
 (b) Correlation function $C(r)$ in two dimensions for the direction parallel to the flow momentum with disorder strength $W=1$ and system size $L=500$ for flow momentum $K=0.1$ , $0.125$, $0.15$, $0.175$, $0.2$, $0.225$, $0.25$ from top to bottom.
 The abscissa and ordinate are shown in log scales.
 (c) Inverse correlation length $l_{\mathrm{c}}^{-1}$ in one dimension for disorder strength $W=0.2$, $0.3$, and $0.5$.
 The dashed curves show the analytical results given by Eq.~(\ref{l_c_RP}).
 (d) Exponent $\eta$ in two dimensions for disorder strength $W=0.5$, $1$, and $2$.
 The dashed curves indicate the analytical results given by Eq.~(\ref{eta_RP}).}
 \label{fig-correlation}
\end{figure}

Next, we consider the case of moderately strong disorder, in which deviations from the perturbative results should be significant.
Figure \ref{fig-correlation} (a) shows the correlation function $C(r)$ in one dimension for several different system sizes.
The disorder average is taken over 100 realizations of the random potential.
Although the finite-size effect is not small, $C(r)$ exhibits an exponential decay in the small-$r$ region, whose width increases with the system size.
Figure \ref{fig-correlation} (c) shows the inverse correlation length $l_{\mathrm{c}}^{-1}$ obtained from $C(r)$ for $L=500$ plotted against the flow momentum $K$ up to the critical flow momentum $K_{\mathrm{c}}$, above which the stationary solution of Eq.~(\ref{stationary_equation_lattice}) does not exist.
The dashed curves in Fig.~\ref{fig-correlation} (c) show the inverse correlation length given by Eq.~(\ref{l_c_RP}).

Figure \ref{fig-correlation} (b) shows the correlation function $C(r)$ in two dimensions for the direction parallel to the flow momentum.
The disorder average is taken over 10 realizations of the random potential.
We have confirmed that, in contrast to the one-dimensional case, the finite-size effect is rather small in two dimensions.
In addition, the power-law decay of $C(r)$ can be observed for $r>5$.
Figure \ref{fig-correlation} (d) shows the exponent $\eta$ obtained from $C(r)$ for $L=100$.
The dashed curves in Fig.~\ref{fig-correlation} (d) show the exponent given by Eq.~(\ref{eta_RP}).
For both one- and two-dimensional cases, the deviations between the numerical and analytical values of $l_{\mathrm{c}}^{-1}$ and $\eta$ increase as the flow momentum $K$ approaches the critical momentum $K_{\mathrm{c}}$.
From Fig.~\ref{fig-correlation}, we conclude that Eq.~(\ref{C_asymptotic}) holds even for a moderately strong disorder, while the inverse correlation length $l_{\mathrm{c}}^{-1}$ in one dimension and the exponent $\eta$ in two dimensions can deviate from Eqs.~(\ref{l_c_RP}) and (\ref{eta_RP}).

\section{Interference and correlation}
\label{sec:interference_and_correlation}

\begin{figure}
 \centering
 \includegraphics[width=0.45\textwidth]{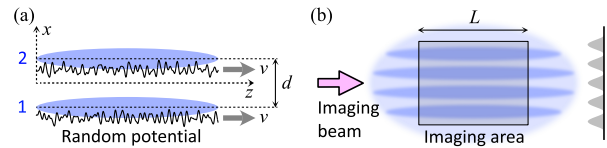}
 \caption{Setup of the interference experiment.
 (a) Two parallel quasi-one-dimensional condensates placed at distance $d$ and subjected to a random potential moving with velocity $v$.
 (b) After free expansion in the transverse direction, an interference pattern is recorded through absorption of an imaging beam.}
 \label{fig-interference}
\end{figure}

We discuss an experimental setup to test our predictions.
The correlation function $C(\mathbf{r})$ can be estimated from the interference pattern between two independent condensates \cite{Bloch-00, Polkovnikov-06, Hadzibabic-06}.
Here, we consider a situation in which two quasi-one-dimensional atomic clouds are placed in parallel at a distance $d$.
Upon these condensates, we impose a random potential moving with velocity $v$ as schematically  illustrated in Fig.~\ref{fig-interference} (a).
This setting is equivalent to an atomic cloud flowing with velocity $-v$ in a random potential at rest.
The random optical potential can be created by a laser beam passing through a diffusive plate \cite{Lye-05, Clement-05, Chen-08, Dries-10}, superposition of incommensurate optical lattices \cite{Tanzi-13}, and holographic imaging with a digital micromirror device \cite{Preiss-15, Zupancic-16}.
Let $z$ be the axial coordinate of the condensates and $x$ be the coordinate along a transverse direction (see Fig.~\ref{fig-interference} (a)).
At the initial time, the trapping potential and the random potential are turned off, and the condensates are allowed to expand freely in the transverse direction.
At time $t$, an interference pattern is recorded on a CCD camera through absorption of an imaging beam directed along the axis of the condensates (see Fig.~\ref{fig-interference} (b)).
The longitudinal length $L$ of an imaging area is determined from the focal length of the imaging beam.

We show that the amplitude of the interference fringes is given by a spatial integral of the correlation function $C(\mathbf{r})$.
Let $\hat{\psi}_1(z)$ and $\hat{\psi}_2(z)$ be the field operators of condensates 1 and 2 at the initial time.
After a time-of-flight of duration $t$, the field operator is given by
\begin{equation}
\hat{\psi}(x,z;t) = \hat{\psi}_1(z) e^{i k_1x} + \hat{\psi}_2(z) e^{i k_2x},
\label{psi_tof}
\end{equation}
where $\hbar k_{1,2}=m(x \pm d/2)/t$ are the momenta of atoms belonging to condensates 1 or 2 at the initial time and detected at position $x$ at time $t$.
The absorption image is given by the density profile integrated along the axial direction, 
\begin{equation}
\hat{\rho}(x) = \int_0^L dz \hat{\psi}^{\dag}(x,z;t) \hat{\psi}(x,z;t).
\end{equation}
From Eq.~(\ref{psi_tof}), the density profile reads 
\begin{equation}
\hat{\rho}(x) = \rho_0 + \hat{A} e^{-ikx} + \hat{A}^{\dag} e^{ikx},
\end{equation}
where $\rho_0$ is a constant background and $k=md/\hbar t$.
The amplitude of the interference fringes is given by
\begin{equation}
\hat{A} = \int_0^L dz \hat{\psi}_1^{\dag}(z) \hat{\psi}_2(z).
\end{equation}
Since the phase of $\hat{A}$ fluctuates randomly for each experimental run, the expectation value of $\hat{A}$ vanishes.
The mean square modulus of $\hat{A}$ is calculated as
\begin{equation}
\langle |\hat{A}|^2 \rangle_{\mathrm{ex}} = L \int_0^L dz \langle \hat{\psi}^{\dag}(z) \hat{\psi}(0) \rangle_{\mathrm{ex}}^2,
\label{A}
\end{equation}
where $\langle ... \rangle_{\mathrm{ex}}$ denotes the average over the experimental runs and we have omitted the indices 1 and 2 in the field operator.

In the classical field theory, the field operator $\hat{\psi}$ in Eq.~(\ref{A}) is replaced by the classical field $\Phi$.
Furthermore, the average over the experimental runs $\langle ... \rangle_{\mathrm{ex}}$ can be interpreted as the disorder average if the random potential is generated independently from shot to shot.
Thus, Eq.~(\ref{A}) is rewritten as
\begin{equation}
\langle |\hat{A}|^2 \rangle_{\mathrm{ex}} = L \int_0^L dz C(z)^2.
\label{A_C}
\end{equation}
From Eq.~(\ref{C_asymptotic}), we have 
\begin{equation}
\bar{A} := \langle |\hat{A}|^2 \rangle_{\mathrm{ex}}^{1/2} \propto \sqrt{l_{\mathrm{c}}L}.
\end{equation}
For the quasi-two-dimensional case, it is convenient to consider the density profile integrated along the additional axis, 
\begin{equation}
\hat{\rho}_{\mathrm{2D}}(x) = \int_0^{L_y} dy \int_0^{L_z} dz \hat{\psi}^{\dag}(x,y,z;t) \hat{\psi}(x,y,z;t).
\end{equation}
Then, we obtain an expression similar to Eq.~(\ref{A_C}) with the integral along the $y$ direction.
From Eq.~(\ref{C_asymptotic}),  for a fixed $L_z \:(\ll L_y)$, we have 
\begin{equation}
\bar{A}_{\mathrm{2D}} \propto L_y^{1-\eta}.
\end{equation}
Thus, by observing how the amplitude of the interference fringes depends on the size of the imaging area, one can estimate $l_{\mathrm{c}}$, $\eta$, and $T_{\mathrm{eff}}$.
In Ref.~\cite{Hadzibabic-06}, this scheme was applied to a two-dimensional trapped Bose gas at thermal equilibrium and a power-law decay of the correlation was observed.

\section{Concluding remarks}
\label{sec:concluding_remarks}

We have demonstrated that the nonequilibrium correlation of the U(1) order parameter of a superflow in a random medium has a one-to-one correspondence to the equilibrium correlation of a clean system at an effective temperature.
What is remarkable about our result is that, whereas there is no inelastic scattering that leads to thermalization, the concept of effective temperature is well-defined.
In particular, the disorder-averaged correlation function is found to exhibits an exponential decay in one dimension and a power-law decay in two dimensions.
We have also proposed an interference experiment of ultracold atomic gases to test our predictions.
It should be noted that the decay of the disorder-averaged correlation function in one and two dimensions can be considered as a nonequilibrium generalization of the Hohenberg--Mermin--Wagner theorem, which predicts the decay of the thermally-averaged correlation function for one- and two-dimensional uniform Bose gases at thermal equilibrium.

The decay of the disorder-averaged correlation does not necessarily imply the breakdown of superflow.
For a fixed realization of a random potential and a given flow velocity below the critical velocity, Eq.~(\ref{General_EOM}) has a unique solution, except for an arbitrary global phase.
The uniqueness of the solution implies that the difference between the phases at any two points has a definite value. 
Thus, the phase coherence of the condensate wave function is not lost, although phases at distant points can have significantly different values.
Since the phase coherence ensures the stability of supercurrent, the decay of the disorder-averaged correlation does not contradict the existence of superfluidity.
However, it has yet to be understood whether the critical velocity remains nonvanishing in the thermodynamic limit.

It is of fundamental importance to investigate the correspondence between the breakdown of superflow at the critical velocity and the equilibrium phase transition to the normal fluid at the critical temperature.
When the flow velocity exceeds the critical velocity, the superflow becomes unstable, and the system undergoes a transition to a turbulent state, which is manifested by the proliferation of vortices.
Here, we recall that the two-dimensional Bose gases exhibit the Berezinskii--Kosterlitz--Thouless (BKT) transition, which is driven by the unbinding of the vortex-antivortex pairs \cite{Hadzibabic-06, Berezinskii-71, Kosterlitz-73, Bishop-78}.
Since the effect of the disorder and flow can be taken into account by an effective temperature, we speculate that the vortex dissociation picture in the BKT transition is also responsible for the breakdown of superflow in a random potential.
Unfortunately, our perturbative approach cannot describe the formation of vortices because we have implicitly assumed that the phase configuration varies slowly in space, and its fluctuations can be described by a quadratic Hamiltonian with respect to the spatial gradient of the phase.
The notion of the effective temperature introduced here may help establish a renormalization group theory for a nonequilibrium BKT transition from coherent to turbulent superflow.

\begin{acknowledgments}
This work was supported by KAKENHI Grant Numbers JP19J00525 and JP18H01145, and a Grant-in-Aid for Scientific Research on Innovative Areas ``Topological Materials Science'' (KAKENHI Grant Number JP15H05855) from the Japan Society for the Promotion of Science.
\end{acknowledgments}

\appendix
\begin{widetext}
\section{Calculation of the mean square relative displacement of the U(1) phase}
\label{appendix:mean_square_relative_displacement}

We calculate the mean square relative displacement of the U(1) phase, $B(\mathbf{r}) = \overline{(\varphi(\mathbf{r}) - \varphi(\mathbf{0}))^2}$.
In terms of the Fourier transform of $\varphi(\mathbf{r})$, $B(\mathbf{r})$ is written as
\begin{equation}
B(\mathbf{r}) = 2 \int \frac{d^D\mathbf{q}}{(2\pi)^D}  \overline{|\varphi_{\mathbf{q}}|^2} (1-\cos \mathbf{q} \cdot \mathbf{r}).
\label{appendix_B_def}
\end{equation}
Here, we have assumed the translation invariance of the disorder-averaged correlation function of $\varphi(\mathbf{r})$ and used the fact that $\varphi(\mathbf{r})$ is real.
In two dimensions, $B(\mathbf{r})$ is not isotropic in the presence of a flow.
We recall that $\overline{|\varphi_{\mathbf{q}}|^2}$ is related to $\overline{|\delta \tilde{n}_{\mathbf{q}}|^2}$ through Eq.~(\ref{phase_fluctuation}).

In one dimension, $B(r)$ is calculated as
\begin{equation}
B(r) = 2 K^2 \bar{n}^{-2} \int_{-\infty}^{\infty} \frac{dq}{2\pi} \overline{|\delta \tilde{n}_{q}|^2} \frac{1-\cos q|r|}{q^2} \simeq 4 K^2 \bar{n}^{-2} \sigma_n^2 \tilde{\xi}_n \int_{0}^{\tilde{\xi}_n^{-1}} \frac{dq}{2\pi} \frac{1-\cos q|r|}{q^2},
\end{equation}
where we have used the fact that $\overline{|\delta \tilde{n}_{q}|^2}$ reduces to $\sigma_n^2 \tilde{\xi}_n$ in the long-wavelength limit $|q| \to 0$ and rapidly vanishes for $|q| > \tilde{\xi}_n^{-1}$.
For $|r| \gg \tilde{\xi}_n$, we have
\begin{equation}
B(r) \simeq 4 K^2 \bar{n}^{-2} \sigma_n^2 \tilde{\xi}_n |r| \int_{0}^{|r|/\tilde{\xi}_n} \frac{dq'}{2\pi} \frac{1-\cos q'}{q'^2} \simeq 4 K^2 \bar{n}^{-2} \sigma_n^2 \tilde{\xi}_n |r| \int_{0}^{\infty} \frac{dq'}{2\pi} \frac{1-\cos q'}{q'^2} = K^2 \bar{n}^{-2} \sigma_n^2 |r|,
\end{equation}
where we have changed the integration variable to $q'=q|r|$.

In two dimensions, $B(\mathbf{r})$ is rewritten as
\begin{eqnarray}
B(\mathbf{r}) &\simeq& 2 \pi^{-2} K^2 \bar{n}^{-2} \sigma_n^2 \tilde{\xi}_n^2 \int_{0}^{\tilde{\xi}_n^{-1}} dq_x \int_{0}^{\tilde{\xi}_n^{-1}} dq_y \frac{q_x^2(1-\cos \mathbf{q} \cdot \mathbf{r})}{(q_x^2+q_y^2)^2},
\label{appendix_B_2D_1}
\end{eqnarray}
where $\mathbf{K}$ is assumed to be parallel to the positive $x$-direction.
First, let us consider the case in which $\mathbf{r}$ is parallel to $\mathbf{K}$.
Then, $B(\mathbf{r})$ is calculated as
\begin{eqnarray}
B(\mathbf{r}) &\simeq& 2 \pi^{-2} K^2 \bar{n}^{-2} \sigma_n^2 \tilde{\xi}_n^2 \int_{0}^{\tilde{\xi}_n^{-1}} dq_x \int_{0}^{\tilde{\xi}_n^{-1}} dq_y \frac{q_x^2(1-\cos q_x |\mathbf{r}|)}{(q_x^2+q_y^2)^2} \nonumber \\
&=& \pi^{-2} K^2 \bar{n}^{-2} \sigma_n^2 \tilde{\xi}_n^2 \int_{0}^{\tilde{\xi}_n^{-1}} dq_x \frac{1-\cos q_x |\mathbf{r}|}{q_x} \left[ \tan^{-1} (\tilde{\xi}_n^{-1} q_x^{-1}) + \frac{\tilde{\xi}_n q_x}{1 + (\tilde{\xi}_n q_x)^2} \right].
\label{appendix_B_2D_2}
\end{eqnarray}
For $|\mathbf{r}| \gg \tilde{\xi}_n$, the dominant contribution to the $q_x$-integral is made from $q_x \simeq |\mathbf{r}|^{-1} \ll \tilde{\xi}_n^{-1}$.
Thus, the quantity in the square brackets in Eq.~(\ref{appendix_B_2D_2}) becomes $\pi/2$ and we have
\begin{equation}
B(\mathbf{r}) \simeq (2\pi)^{-1} K^2 \bar{n}^{-2} \sigma_n^2 \tilde{\xi}_n^2 \int_{0}^{\tilde{\xi}_n^{-1}} dq_x \frac{1-\cos q_x |\mathbf{r}|}{q_x} \simeq (2\pi)^{-1} K^2 \bar{n}^{-2} \sigma_n^2 \tilde{\xi}_n^2 \ln(|\mathbf{r}|/\tilde{\xi}_n).
\end{equation}
In a similar manner, if $\mathbf{r}$ is perpendicular to $\mathbf{K}$, $B(\mathbf{r})$ is calculated as
\begin{eqnarray}
B(\mathbf{r}) &\simeq& 2 \pi^{-2} K^2 \bar{n}^{-2} \sigma_n^2 \tilde{\xi}_n^2 \int_{0}^{\tilde{\xi}_n^{-1}} dq_x \int_{0}^{\tilde{\xi}_n^{-1}} dq_y \frac{q_x^2(1-\cos q_y |\mathbf{r}|)}{(q_x^2+q_y^2)^2} \nonumber \\
&=& \pi^{-2} K^2 \bar{n}^{-2} \sigma_n^2 \tilde{\xi}_n^2 \int_{0}^{\tilde{\xi}_n^{-1}} dq_y \frac{1-\cos q_y |\mathbf{r}|}{q_y} \left[ \tan^{-1} (\tilde{\xi}_n^{-1} q_y^{-1}) - \frac{\tilde{\xi}_n q_y}{1 + (\tilde{\xi}_n q_y)^2} \right] \nonumber \\
&\simeq& (2\pi)^{-1} K^2 \bar{n}^{-2} \sigma_n^2 \tilde{\xi}_n^2 \ln(|\mathbf{r}|/\tilde{\xi}_n).
\end{eqnarray}
Thus, the long-distance behavior of $B(\mathbf{r})$ is isotropic.
\end{widetext}

\end{document}